\newcommand{\CLASSINPUTtoptextmargin}{0.8in}
\newcommand{\CLASSINPUTbottomtextmargin}{0.8in}
\documentclass[conference]{IEEEtran}
\IEEEoverridecommandlockouts
\usepackage{cite}
\usepackage{amsmath,amssymb,amsfonts}
\usepackage{algorithmic}
\usepackage{graphicx}
\usepackage{textcomp}
\usepackage{lettrine}
\usepackage{xcolor}
\usepackage{booktabs}
\usepackage{subcaption}
\usepackage{microtype}

\def\BibTeX{{\rm B\kern-.05em{\sc i\kern-.025em b}\kern-.08em
    T\kern-.1667em\lower.7ex\hbox{E}\kern-.125emX}}
\begin{document}

\title{Harnessing Flexible Spatial and Temporal Data Center Workloads for Grid Regulation Services \\
}

\author{\IEEEauthorblockN{Yingrui Fan}
\IEEEauthorblockA{\textit{Thayer School of Engineering} \\
\textit{Dartmouth College}\\
Hanover, NH, USA \\
Yingrui.fan.th@dartmouth.edu}
\and
\IEEEauthorblockN{Junbo Zhao}
\IEEEauthorblockA{\textit{Thayer School of Engineering} \\
\textit{Dartmouth College}\\
Hanover, NH, USA \\
Junbo.zhao@dartmouth.edu}
\and
}
\IEEEaftertitletext{\vspace{-3em}}
\IEEEsetheadermargin{t}{0.5in}
\IEEEsetfootermargin{b}{0.5in}
\maketitle

\begin{abstract}
Data centers (DCs) are increasingly recognized as flexible loads that can support grid frequency regulation. Yet, most existing methods treat workload scheduling and regulation capacity bidding separately, overlooking how queueing dynamics and spatial–temporal dispatch decisions affect the ability to sustain real-time regulation. As a result, the committed regulation may become infeasible or short-lived. To address this issue, we propose a unified day-ahead co-optimization framework that jointly decides workload distribution across geographically distributed DCs and regulation capacity commitments. We construct a space–time network model to capture workload migration costs, latency requirements, and heterogeneous resource limits. To ensure that the committed regulation remains deliverable, we introduce chance constraints on instantaneous power flexibility based on interactive load forecasts, and apply Value-at-Risk queue-state constraints to maintain sustainable response under cumulative regulation signals. Case studies on a modified IEEE 68-bus system using real data center traces show that the proposed framework lowers system operating costs, enables more viable regulation capacity, and achieves better revenue–risk trade-offs compared to strategies that optimize scheduling and regulation independently.
\end{abstract}


\section{Introduction}
The rapid growth of artificial intelligence (AI), cloud computing, and large-scale internet services has led to a substantial increase in the electricity consumption of DCs. By 2025, global DC power demand is projected to approach 2\% of total electricity usage \cite{iea2025}. Unlike conventional inflexible loads, DCs host workload types that often allow controlled temporal scheduling and geographical workload migration. Such inherent flexibility considers DCs as promising resources for grid support services, including frequency regulation \cite{alkez2021}. Recently, the California Energy Commission has begun recognizing DCs as dispatchable load resources with the potential to provide services \cite{Fan2025}.

However, existing work hasn't fully unlocked this potential. Initial studies have demonstrated the ability of DCs to modulate workloads in response to grid conditions \cite{dvorkin2025}. Many of these approaches optimize DC power usage in isolation and do not jointly consider system-level constraints such as transmission congestion \cite{ding2025}. As a result, these strategies may improve DC operational efficiency but provide only marginal benefits to overall grid flexibility. Meanwhile, market-participation models that enable DCs to offer frequency regulation capacity often assume simplified latency and workload behavior \cite{zou2025}. Such models overlook the fact that DC frequency response is inherently constrained by the evolution of workload queues and service-level guarantees. When downward regulation slows computing, tasks accumulate; when upward regulation accelerates computing, the backlog depletes. Sustained imbalance prevents the DC from maintaining its committed regulation signal, leading to non-compliant response \cite{ren2025}.

Several papers have sought to artificially increase regulation flexibility. For example, virtual workload injection can expand regulation headroom but raises energy consumption and operational cost \cite{wang2019}. Similarly, cryptocurrency mining operations can provide large regulation swings due to highly elastic workloads \cite{menati2023}. However, these approaches are not directly applicable to modern cloud DCs, which must maintain strict service-level agreements and handle heterogeneous workloads with differentiated performance sensitivities.


To address this gap, this paper proposes a day-ahead spatio-temporal co-optimization framework that
jointly determines multi-data-center workload scheduling and frequency regulation capacity 
commitments. Day-ahead workload placement shapes baseline power and computation buffers, defining feasible regulation capacity. Workload migration is modeled as flows on a unified space–time network with latency and hardware constraints. Feasibility is evaluated using historical regulation 
signals, chance constraints limit instantaneous responses, and VAR constraints capture sustained energy deviations, ensuring probabilistic deliverability of the committed capacity.


 
\section{Proposed Framework}

Modern DCs are modeled as geographically distributed cyber–physical systems where workloads are flexibly dispatched across space and time. To capture this, we introduce a space–time network representation that maps each DC at each time slot into a virtual node within a 2D grid structure $(l,t)$, subsequently indexed as $i$ through a mapping function.
\begin{equation}
p(l,t) = (l-1)T + t, \quad \forall l \in \mathcal{L}, \forall t \in \mathcal{T}
\end{equation}
Here, $p(l,t)\in[1,\psi]$ identifies the virtual node corresponding to DC $l$ at time $t$, and $\psi = N\times T$ represents the total number of virtual nodes in the system.  
This maps all nodes onto a single index dimension, which simplifies subsequent matrix operations and makes space–time interactions more direct.

\begin{figure}[htbp]
\vspace{-5pt}
    \hspace*{0cm} 
    \includegraphics[width=0.5\textwidth]{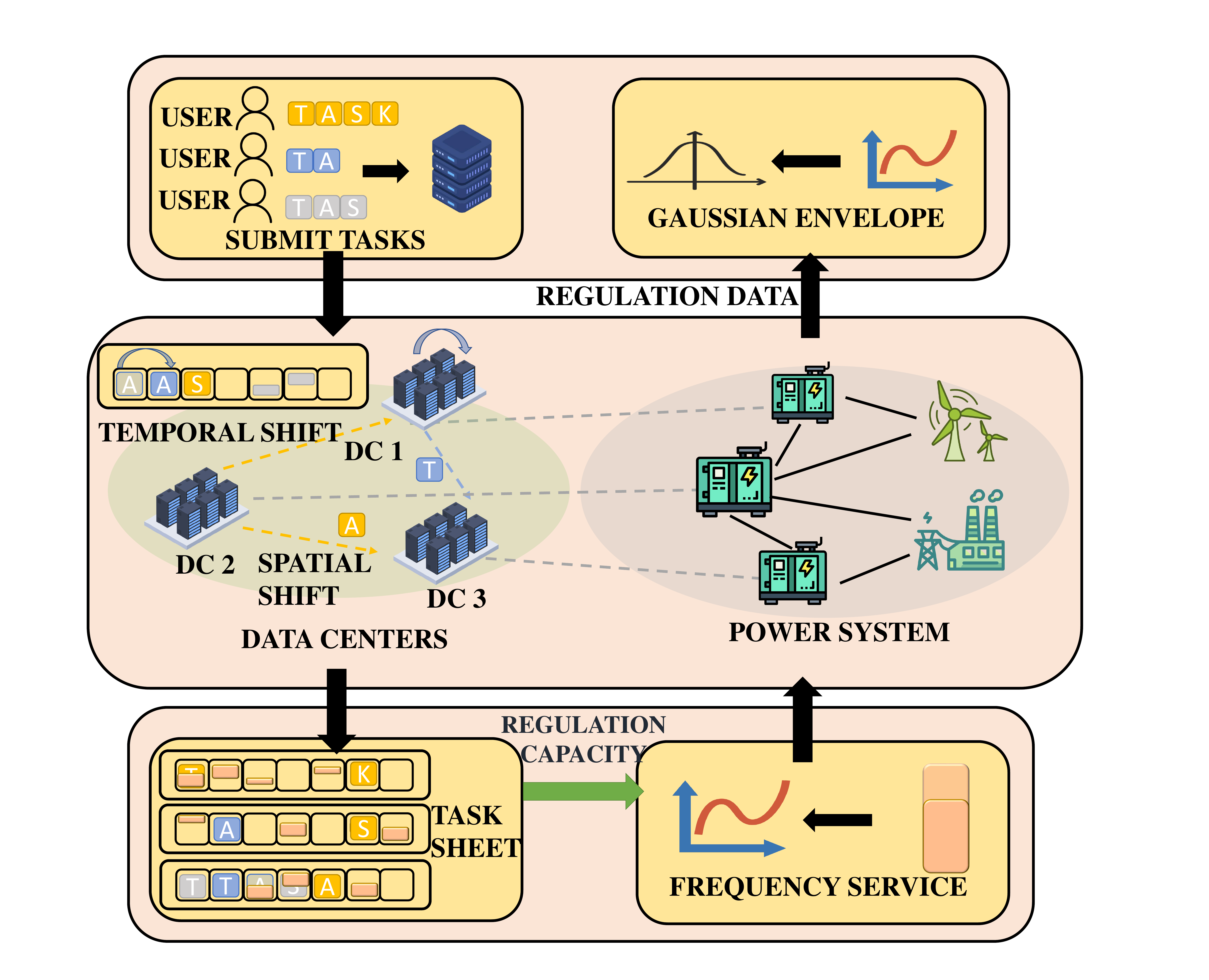}
    \caption{Illustration for co-optimization model.}
    \label{frame}
\end{figure}

Rather than modeling scheduling at the individual job level, we use an aggregated execution-rate representation. Let $x_{i,t,l} \in [0,1]$ denote the fraction of workload $i$ processed at data center $l$ during time slot $t$.
To couple computing workloads with physical power consumption, we define $d_i$ as the effective load contribution of job $i$ when it is scheduled.  
The aggregate load vector $\boldsymbol{\bar\vartheta}\in\mathbb{R}^{\psi}$, representing the total power demand at each space–time node, can thus be expressed as:
\begin{equation}
\overline{\vartheta}_{p(l,t)} = \sum_{i} d_{i} x_{i,t,l}, \quad \forall l \in \mathcal{L}, \forall t \in \mathcal{T}
\end{equation}
This formulation bridges the cyber scheduling layer and the physical electricity layer, such that changes in workload allocation directly affect the spatial and temporal power profiles of the data center network.

This paper adopts an undirected-edge-with-fixed-orientation convention, meaning that each link is represented once, and the transfer direction is indicated by the sign of the corresponding flow variable.  
The number of spatial and temporal virtual links is given by:
\begin{equation}
K_{\text{sp}} = \binom{N}{2}\,T, \qquad K_{\text{tm}} = N\,(T-1),
\end{equation}
where $\binom{N}{2}$ represents the number of possible DC pairs, while the temporal term $N(T-1)$ captures adjacent time-slot connections for each DC. So the total number of virtual links in the network is $K = K_{\text{sp}} + K_{\text{tm}}$.


To consider a set of users indexed by $u$, and each user submits computing jobs that must be executed at one of the data centers $l \in \mathcal{L}$. Let $d(u,l)$ denote the geographical distance-dependent latency cost between user $u$ and data center $l$. In the context of this research, Quality of Service (QoS) is specifically quantified as the geographical latency between the user and the data center. The effective QoS latency at time $t$ is the weighted average latency experienced by all scheduled computing jobs:
\begin{equation}
    \mathcal{L}_{t} = \frac{\sum_{i \in \mathcal{J}} \sum_{l \in \mathcal{L}} d(u(i),l) \, x_{i,t,l}}{\sum_{i \in \mathcal{J}} \sum_{l \in \mathcal{L}} x_{i,t,l}},
    \label{eq:latency}
\end{equation}
where $\mathcal{J}$ denotes the set of all computing jobs.

To enforce QoS guarantees, we compare the latency under shifting with the baseline assignment. The baseline latency $\overline{\mathcal{L}}_t$ is pre-calculated using Equation (4) based on the nominal workload assignment $x_{i,t,l}^{base}$ that satisfies basic operational requirements without considering grid flexibility services. The allowable latency deviation is upper bounded by a tolerance factor $\Delta_{\text{QoS}}$, producing the QoS constraint:
\begin{equation}
    \mathcal{L}_{t} - \overline{\mathcal{L}}_{t} \le \Delta_{QoS}, \quad \forall t \in \mathcal{T}
    \label{eq:latency-constraint}
\end{equation}

Constraint~\eqref{eq:latency-constraint} ensures that workload shifting decisions do not increase end-user delay beyond a controllable and system-specified tolerance threshold.

The scheduling decision $x_{i,t,l}$ must also satisfy the physical hardware limitations within each data center. At any time $t$, the total computing resources consumed by all computing tasks $i$ ($i=1, ..., M$) deployed on data center $l$ cannot exceed the available capacity of that data center. This set of local capacity constraints is represented as follows:
\begin{subequations}
\begin{align}
\sum_{i=1}^M{x_{i,t,l}}\cdot R_{i}^{\mathrm{CPU}}\le CPU_{l,t}^{\mathrm{cap}}\qquad \quad \forall l \in \mathcal{L}, \forall t \in \mathcal{T}
\\
\sum_{i=1}^M{x_{i,t,l}}\cdot R_{i}^{\mathrm{MEM}}\le MEM_{l,t}^{\mathrm{cap}}\qquad \quad \forall l \in \mathcal{L}, \forall t \in \mathcal{T}
\\
\sum_{i=1}^M{x_{i,t,l}}\cdot R_{i}^{\mathrm{I}/\mathrm{O}}\le I/O_{l,t}^{\mathrm{cap}}\qquad \quad \forall l \in \mathcal{L}, \forall t \in \mathcal{T}
\end{align}
\end{subequations}
where $R_{i}^{\text{CPU}}$, $R_{i}^{\text{MEM}}$, and $R_{i}^{\text{I/O}}$ represent the standardized amount of resources required to execute job $i$. $CPU_{l,t}^{\text{cap}}$, $MEM_{l,t}^{\text{cap}}$, and $I/O_{l,t}^{\text{cap}}$ are the maximum available physical resource capacities at DC $l$ at time $t$. These constraints ensure that while workload shifting occurs at the network level, the underlying task scheduling remains feasible.
\section{FEASIBLE FREQUENCY REGULATION CAPACITY MODELING}

\subsection{Probabilistic Model}

To participate in the frequency regulation, each DC $l$ must commit to this baseline load while also providing a feasible symmetric regulation capacity, denoted $R_{l,t}$. However, real-time delivery of $R_{l,t}$ faces instantaneous power uncertainty and cumulative energy uncertainty.

\subsubsection{Instantaneous Power Constraints using Gaussian Envelope}

The actual instantaneous power $P_{l,t,k}$ must remain within the data center's physical operating bounds $[P_{l,t}^{\text{min}}, P_{l,t}^{\text{max}}]$, where $k$ denotes the intra-slot sampling index within each day-ahead time period $t$, capturing the high-frequency fluctuations of the regulation signal $s_k$. The total power is the difference of the regulated baseline and the stochastic interactive load: $P_{l,t,k} = \overline{\vartheta}_{p(l,t)} - s_k \cdot R_{l,t}$.

This creates two-sided chance constraints. The downward regulation limit is constrained by maximum physical power $P_{l,t}^{\text{max}}$:
\begin{equation}
\overline{\vartheta}_{p(l,t)} + R_{l,t} \le P_{l,t}^{max}, \quad \forall l \in \mathcal{L}, \forall t \in \mathcal{T}
\end{equation}

The upward regulation limit is constrained by the combined stochasticity of the signal and the interactive load. This joint chance constraint must be met with reliability $1-\epsilon_p$:
\begin{equation}
\mathbb{P}\{\overline{\vartheta}_{p(l,t)} - s_{k} \cdot R_{l,t} \ge P_{l,t}^{min}\} \ge 1-\epsilon_{p}, \quad \forall l \in \mathcal{L}, \forall t \in \mathcal{T}
\end{equation}

Downward regulation increases power consumption towards the hardware limit $P^{max}$, which expands computational headroom and typically does not harm task efficiency. Conversely, upward regulation reduces power, creating a direct conflict with the data center's internal workloads. Thus, while (7) represents a deterministic hardware bound, chance constraint (8) is applied to upward regulation to maximize profit by balancing market revenue against the risk of service degradation.
A significant challenge arises because historical analysis shows the regulation signal $s_k$ is non-Gaussian. Directly using an empirical distribution or a standard Gaussian approximation in a chance-constrained optimization leads to either intractable problems or an underestimation of tail risk.

To handle non-Gaussian signals, we adopt the Gaussian Envelope method to construct a conservative Gaussian outer approximation whose CDF upper-bounds the empirical CDF of historical regulation data. This ensures probabilistic feasibility even under worst-case tail fluctuations. The reformulated SOCP constraint is:
\begin{equation}
\boldsymbol{\alpha}^\top \boldsymbol{\mu_J} + \Phi^{-1}(1-\epsilon_p) \|\boldsymbol{\Sigma_J}^{1/2}\boldsymbol{\alpha}\|_2 \le \beta
\end{equation}
where $\boldsymbol{\alpha} = [R_{l,t}]^\top$. $\boldsymbol{\mu_J}$ and $\boldsymbol{\Sigma_J}$ are the mean and covariance of the outer-approximated distribution derived from empirical traces. $\beta = \overline{\vartheta}_{p(l,t)} - P_{l,t}^{\text{min}}$ is the deterministic baseline power margin. This formulation ensures the optimization remains tractable, convex, and robust against non-Gaussian risks.

\subsubsection{Cumulative Energy Constraints using VaR}



The queue state $Q_{l,t}$ represents the amount of deferrable computing workload remaining at data center $l$ at time $t$. This queue evolves according to standard workload conservation dynamics. While instantaneous power constraints (7)-(8) address high-frequency signal fluctuations, the queue state captures cumulative energy deviations over time. We apply VaR to these cumulative constraints to bound sustained imbalances using historical data, ensuring long-term service stability with linear tractability. Let $A_{l,t}$ denote the newly arrived deferrable workload assigned to data center $l$ at time $t$, and let $S_{l,t}$ denote the amount of workload served at $l$ at time $t$. The queue dynamics are given by:
\begin{equation}
    Q_{l,t+1} = Q_{l,t} + A_{l,t} - S_{l,t}.
\end{equation}

Based on these scheduling decisions, the baseline queue trajectory without regulation over horizon $\tau$ is defined as:
\begin{equation}
    Q_{l,\tau}^{\text{base}} = Q_{l,0} + \sum_{t=1}^{\tau} \left( A_{l,t} - \bar \vartheta_{p(l,t)} \right).
\end{equation}

This equation establishes the explicit coupling between the space--time workload scheduling model in Section II and the regulation feasibility constraints in Section III. Real-time frequency regulation achieves the effective computing rate by modifying the instantaneous power consumption, and thus the cumulative regulation energy $R_{l,t} \cdot \tilde{S}_{\tau}$ shifts the queue state away from its baseline value $Q_{l,\tau}^{\text{base}}$, where $\tilde{S}_{\tau}$ is defined as the stochastic cumulative regulation signal, calculated as the sum of the regulation signal $s_k$ over the sub-hour interval $\tau$ multiplied by the sampling interval $\Delta k$.

We analyze the queue state $Q_{l,\tau}$ over $\tau$, ensuring it remains within physical bounds $[Q_{l}^{\text{min}}, Q_{l}^{\text{max}}]$. The state is impacted by the uncertain cumulative regulation energy $\tilde{E}_{\text{reg},\tau}=R_{l,t}\cdot\tilde{S}_{\tau}$.
The actual state is $Q_{l,\tau}^{\text{base}}+\tilde{E}_{\text{reg},\tau}$. This forms a two-sided chance constraint:
\begin{equation}
\mathbb{P}\{Q_{l}^{\text{min}}\le Q_{l,\tau}^{\text{base}}+R_{l,t}\cdot\tilde{S}_{\tau}\le Q_{l}^{\text{max}}\}\ge1-\epsilon_{e}
\end{equation}

For this cumulative constraint, this paper adopt the VaR method, using the $\epsilon_{e}$ and $1-\epsilon_{e}$ quantiles of $\tilde{S}_{\tau}$ derived from historical data:
\begin{align}
S_{\tau}^{\text{low}} &= \text{VaR}_{\epsilon_{e}}(\tilde{S}_{\tau}) \\
S_{\tau}^{\text{high}} &= \text{VaR}_{1-\epsilon_{e}}(\tilde{S}_{\tau})
\end{align}

The chance constraint is then robustly reformulated as two linear constraints:
\begin{align}
Q_{l,\tau}^{base} + R_{l,t} \cdot S_{\tau}^{high} \le Q_{l}^{max}, \quad \forall l \in \mathcal{L}, \forall t \in \mathcal{T}, \forall \tau \in \mathcal{T}_{\tau}  \\
Q_{l,\tau}^{base} + R_{l,t} \cdot S_{\tau}^{low} \ge Q_{l}^{min}, \quad \forall l \in \mathcal{L}, \forall t \in \mathcal{T}, \forall \tau \in \mathcal{T}_{\tau}
\end{align}

Constraint~(15) prevents queue overflow from sustained upward regulation, while (16) prevents queue starvation from sustained downward regulation in sub intervals $\mathcal{T}_{\tau}$.

\subsection{Optimization Model}
The proposed co-optimization framework aims to minimize the total net system cost, subject to a set of technical constraints governing both the power grid and the data center network. The objective function and the associated operational constraints are detailed below. Unless otherwise specified, all constraints in this section apply to DC nodes $l \in \mathcal{L}$, time slots $t \in T$, generation nodes $g \in \mathcal{G}$, and all buses $b \in \mathcal{B}$:
{
\allowdisplaybreaks
\begin{subequations} \label{eq:full_model}
\begin{align}
&\text{minimize} \quad 
 \sum_{t=1}^{T} \Bigg(
    \sum_{g \in \mathcal{G}} C_g(p_{g,t})
    + \sum_{b \in \mathcal{B}} C^{\text{penal}} q_{b,t}  \nonumber \\
 &\qquad\qquad
    - \sum_{l \in \mathcal{L}} (c_{rc,t} + c_{rp,t} \cdot \bar{m}_t) R_{l,t}
\Bigg)
\label{eq:objective} \\[6pt]
\text{subject to:} \nonumber \\[2pt]
\begin{split}\sum_{g \in \mathcal{G}_b} p_{g,t} - \sum_{l \in \mathcal{L}_{b}} \bar{\vartheta}_{p(l,t)} - p_{b,t}^{base} \\ + q_{b,t} = \sum_{e=(b,j) \in \mathcal{E}} B_k (\theta_{b,t} - \theta_{j,t}) \end{split} \label{eq:power_balance}\\
&- \bar{f}_k \le \, B_k (\theta_{b,t} - \theta_{j,t}) \le \bar{f}_k \label{eq:line_limits} \\
&\underline{p}_g u_{g,t} \le \, p_{g,t} \le \bar{p}_g u_{g,t}, \label{eq:gen_limits} \\
&p_{g,t} - p_{g,t-1} \le \, \overline{p}_g^{\uparrow} u_{g,t-1} + \overline{p}_{\text{up}} (u_{g,t} - u_{g,t-1}) \label{eq:ramp_up} \\
&p_{g,t-1} - p_{g,t} \le \, \overline{p}_g^{\downarrow} u_{g,t} + \overline{p}_{\text{dw}} (u_{g,t-1} - u_{g,t}) \label{eq:ramp_down} \\
&\sum_{t \in \mathcal{T}} \sum_{l \in \mathcal{L}} x_{i,t,l} = 1, \quad \forall i \in \mathcal{J}\label{total}
\end{align}
\text{Equations (5), (6), (7), (8), (12), (15) and (16).}
\end{subequations}
}

The objective function consists of the power consumption cost and the penalty for unfinished tasks, offset by the revenue from frequency regulation services. Specifically, $C_g$ denotes the electricity price determined by LMPs, $p_{g,t}$ represents the power output of generator $g$ at time $t$, and $C^{penal}$ is a sufficiently large penalty coefficient associated with load shedding $q_{b,t}$ at bus $b$. The parameters $c_{rc,t}$ and $c_{rp,t}$ define the regulation payment components, while $\bar{m}_t$ reflects the regulation signal amplitude or the performance score.Regarding the constraints, \eqref{eq:power_balance} describes the nodal power balance at bus $b$ and time $t$, where $p_{b,t}^{base}$ accounts for the baseline load consumption. Transmission line limits are enforced in \eqref{eq:line_limits}, which restricts the power flow, calculated via the DC power flow approximation $B_k (\theta_{b,t} - \theta_{j,t})$—within the maximum rated capacity $\bar{f}_k$. Generator operational limits, including capacity bounds and ramping constraints, are specified in \eqref{eq:gen_limits}, \eqref{eq:ramp_up}, and \eqref{eq:ramp_down}, respectively. Finally, constraint \eqref{total} ensures that all scheduled computing tasks are fully completed within the optimization horizon.

\section{Case Studies}
Simulations are implemented on a hardware platform of AMD Ryzen 9 5900X 12-CoreCPU; in this study, we conduct a case study involving eight geo-distributed data centers to evaluate the effectiveness of the proposed coordinated bidding strategy based on the modified IEEE 68-bus system. DCs are on buses \{17, 29, 32, 49, 50, 54, 40, 66\}. The software system for the current optimization is Julia 1.11.5, implemented based on the commercial Gurobi solver. Power demand for each clustered workload $d_i$ is designed to be 1.7 kWh/task, referring to the LLM-powered Google search \cite{devries2023}. We obtain regulation prices and RegD signal from the PJM market. The workload composition is structured with 50\% fixed load. 30\% of interactive tasks, and 20\% of deferrable tasks. The computing job trace dataset is sourced from Alibaba data center \cite{alibaba_clusterdata}. All load data are from the 2018 ISO-NE and NYISO  \cite{iso-ne_markets} \cite{nyiso_data}.
\begin{figure}[htbp]
    \centering
    \begin{subfigure}[b]{0.23\textwidth}
        \centering
        \includegraphics[width=\linewidth]{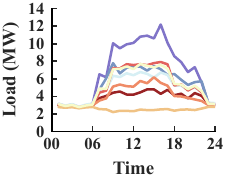}
        \caption{No coordination}
        \label{traditional}
    \end{subfigure}\hfill
    \begin{subfigure}[b]{0.23\textwidth}
        \centering
        \includegraphics[width=\linewidth]{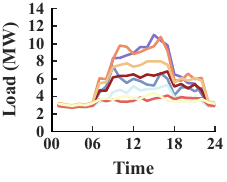}
        \caption{Allow spatial shifting}
        \label{spatial}
    \end{subfigure}

    \medskip

    \begin{subfigure}[b]{0.23\textwidth}
        \centering
        \includegraphics[width=\linewidth]{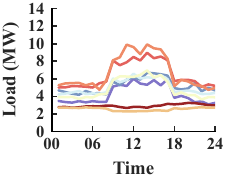}
        \caption{Allow temporal shifting}
        \label{temporal}
    \end{subfigure}\hfill
    \begin{subfigure}[b]{0.23\textwidth}
        \centering
        \includegraphics[width=\linewidth]{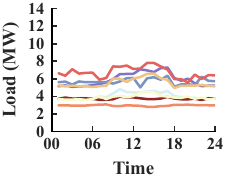}
        \caption{Joint method}
        \label{joint}
    \end{subfigure}
    
    \vspace{3pt} 
    \includegraphics[width=0.45\textwidth]{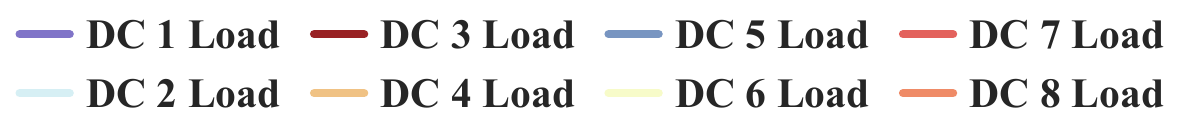}
    
    \caption{The load across the testing dataset under different coordination models. The curve defines the final cost and frequency regulation capacity.}
    \label{load}
\end{figure}

\vspace{-0.2cm}
\subsection{Effectiveness of Load Shifting}\label{AA}
Fig. \ref{load} compares the system load profiles under different coordination strategies.
With spatial workload migration shown in Fig. \ref{spatial}, workloads are shifted to lower-cost regions like DC 8, reducing local congestion and operating costs, while the overall system load curve remains unchanged.
With temporal shifting shown in Fig. \ref{temporal}, flexible tasks are deferred to lower-cost periods, resulting in a flatter load curve and reduced peak demand, which benefits grid stability.
The joint spatial–temporal strategy in Fig. \ref{joint} further balances load across both space and time, simultaneously relieving nodal stress and lowering system peaks. This integrated approach achieves the most cost-efficient and grid-friendly operational outcome.

\begin{figure}[htbp]
    \centering
    \includegraphics[width=0.4\textwidth]{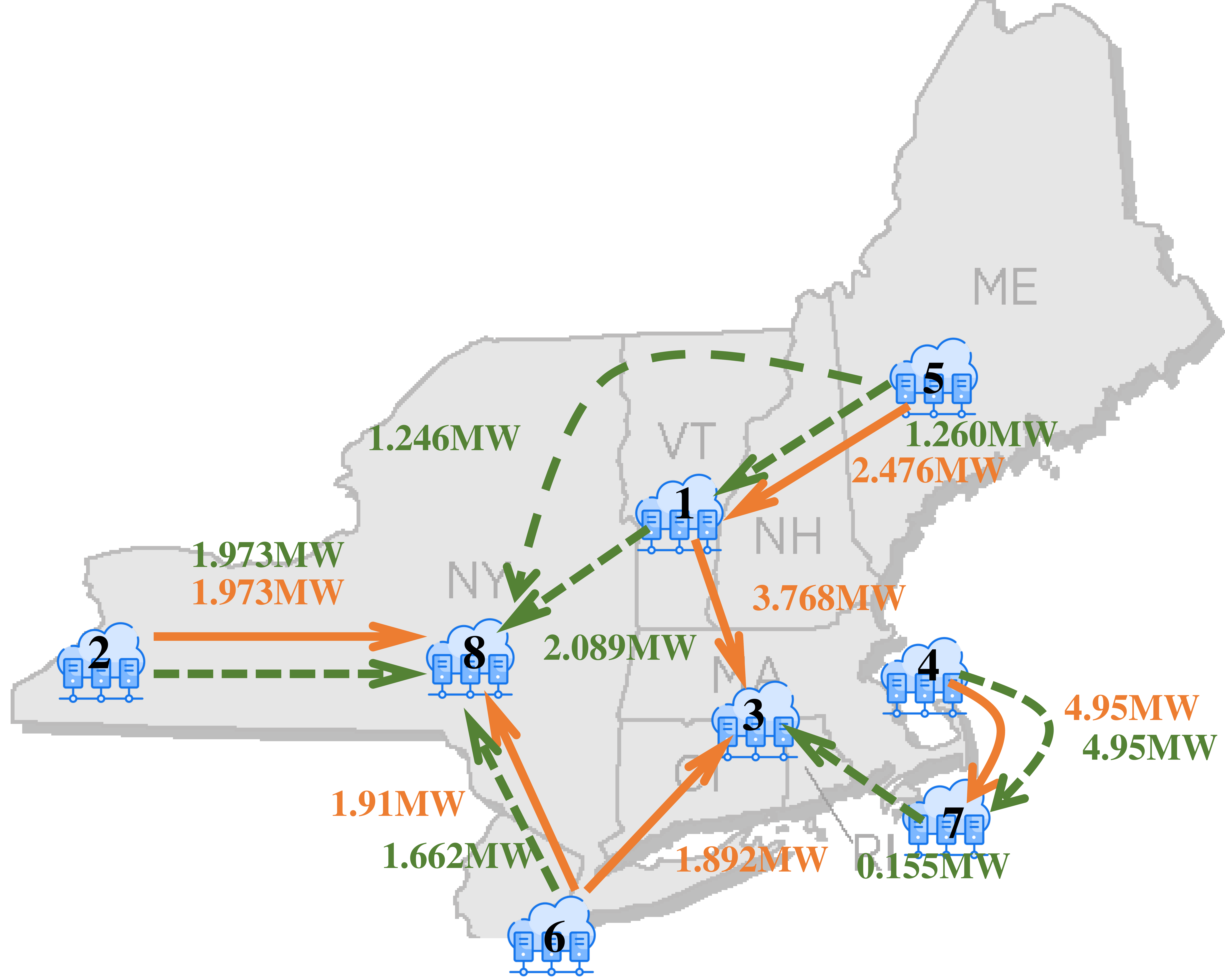}
    \caption{68-bus NY-NE ISO system with 8 data centers. The arrows show active virtual links between different DCs, real-time coordination under different coordination solutions for 100\% latency loss and 25\% latency loss, as green links and orange links.}
    \label{geodispatch}
\end{figure}

The geo-coordinated workload dispatch strategy is illustrated in Fig. \ref{geodispatch}. Under conditions of relaxed latency constraints, workloads are flexibly migrated across geographically distributed DCs to exploit spatial cost differentials. In this case, workloads originating from DC 5 can be economically redirected to remote DCs such as DC 8, where the electricity price is lower. 

Conversely, when latency tolerance becomes more stringent, the spatial flexibility of workload placement is significantly constrained. Tasks must then be assigned to nearby facilities, such as DC 1, in order to satisfy latency limits. This restricted coordination reduces the opportunity to exploit geographical price differences, ultimately resulting in less favorable economic performance.

\begin{table}[h]
\centering
\setlength{\tabcolsep}{3pt} 
\renewcommand{\arraystretch}{0.9} 
\caption{Comparison of DCs in different models}
\begin{tabular}{lccccc}
\toprule
Model & \begin{tabular}[c]{@{}c@{}}Average\\ Load (MW)\end{tabular} 
& \begin{tabular}[c]{@{}c@{}}Total \\ Cost (k\$)\end{tabular}
& \begin{tabular}[c]{@{}c@{}}Average\\ Regulation\\ Capacity (MW)\end{tabular}
& \begin{tabular}[c]{@{}c@{}}Regulation\\ Profit (k\$)\end{tabular}
& \begin{tabular}[c]{@{}c@{}}Net\\ Cost (k\$)\end{tabular} \\ 
\midrule
Decoupled     & 38.30 & 142.50 & 4.50  & 45.89  & 96.61 \\
Independent   & 38.25 & 140.65 & 10.23 & 103.98 & 36.67 \\
Cooperative   & 38.24 & 138.84 & 12.84 & 130.95 & 7.89  \\
\bottomrule
\end{tabular}
\label{comparison}
\end{table}
To validate the effectiveness of the proposed framework, we compared three strategies: 1) a decoupled model, which uses a sequential optimization approach of scheduling followed by adjustment; 2) an independent model, which considers adjustment but lacks cross-center coordination; and 3) a collaborative model, which is the spatio-temporal joint optimization framework proposed in this paper.

In Table \ref{comparison}, the decoupled strategy performs the worst because Phase 1 fixes the workload allocation, leaving minimal flexibility to supply scalable regulation capacity in Phase 2. The independent strategy, although capable of proactively providing considerable scalable capacity, fails to minimize total system power cost due to the absence of coordination across data centers. In contrast, the proposed approach achieves global coordination: it strategically shifts workloads toward nodes with lower marginal system cost while simultaneously constructing the maximum feasible scalable capacity. This integrated optimization leads to the lowest overall net cost.
\subsection{Frequency Regulation Performance}
\begin{figure}[htbp]
\centering
\begin{subfigure}[t]{0.24\textwidth}
    \centering
    \includegraphics[width=\textwidth]{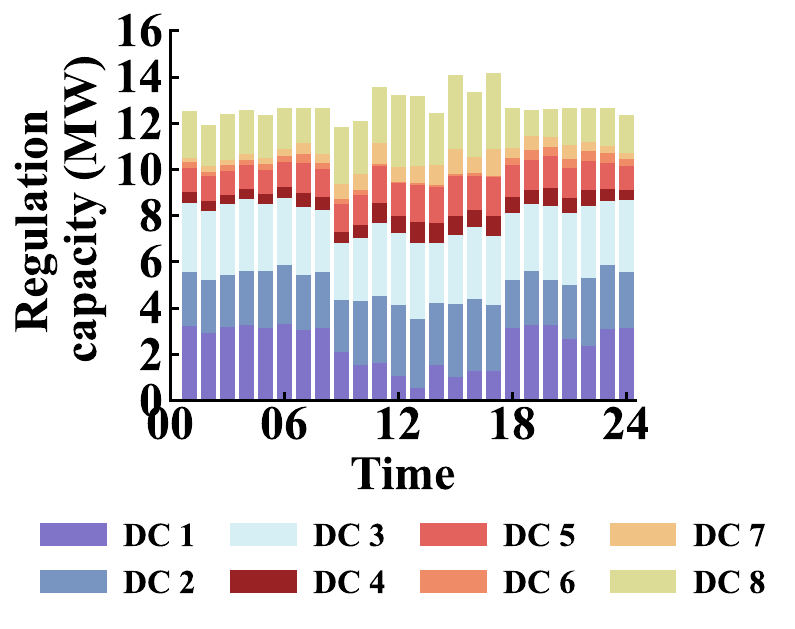}
    \caption{Regulation capacity across data centers}
    \label{capacity}
\end{subfigure}
\hfill
\begin{subfigure}[t]{0.24\textwidth}
    \centering
    \includegraphics[width=\textwidth]{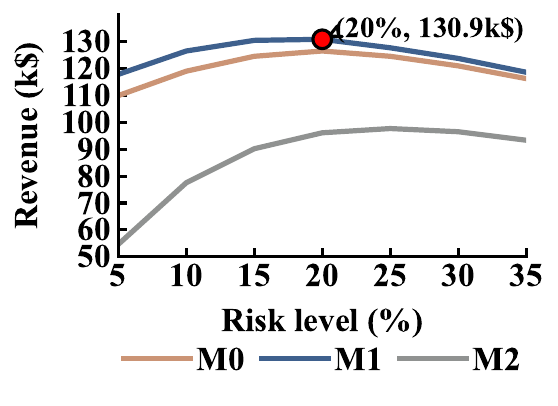}
    \caption{Revenue–risk tradeoff}
    \label{subfig:tradeoff}
\end{subfigure}
\caption{Regulation capacity distribution and revenue–risk tradeoff under different signal modeling approaches.}
\label{fig:combined}
\end{figure}

As shown in Fig. \ref{capacity}, for low-power data centers such as DC4, the narrow gap between the upper and lower operating limits constrains the adjustable range, resulting in a relatively small regulation capacity $R$. Conversely, when the baseline power consumption is excessively high---as in the case of DC1 during early morning hours---the upper limit becomes binding, again reducing the feasible range of $R$. These observations highlight that both insufficient and excessive baseline utilization can impair a data center’s regulation capability, indicating the importance of maintaining balanced operating levels.

To address the non-Gaussian characteristics of real regulation signals, we compare direct Gaussian distribution, Gaussian envelope approximation and Wasserstein-DRO \cite{ren2025}. Among these, M1 achieves the highest revenue under PJM's allowable 25\% non-compliance rate by maximizing the realizable regulation capacity. This performance stems from the Gaussian envelope approximation, which expands the feasible signal region to avoid underestimating tail risks, thereby ensuring that the optimization remains robust and feasible even when the actual regulation signal deviates from Gaussian assumptions.

The proposed framework balances economic performance and operational reliability by leveraging the Gaussian envelope to manage non-Gaussian signal risks without excessive conservativeness. The integration of VaR-based constraints (14)-(15) ensures sustainable energy balance over long horizons while maintaining linear tractability. Furthermore, the space-time network representation limits the optimization complexity to a polynomial scale relative to $N \times T$, facilitating its application in larger grid configurations. The unified co-optimization mechanism remains inherently adaptive to varying levels of workload flexibility through the dynamic modulation of $x_{i,t,l}$.

\section{Conclusion}
This work develops a spatial--temporal co-optimization framework that enables DCs to provide day-ahead frequency regulation while respecting latency and operational constraints. By coordinating workload shifting across both locations and time, and by applying a Gaussian envelope to handle non-Gaussian regulation signals, the framework increases feasible regulation capacity and improves market revenue. Case studies on the modified 68-bus system show that spatial shifting alleviates nodal stress, temporal shifting smooths system peaks, reduces operation costs, and their joint use achieves the best overall performance.

\vspace{12pt}


\begin{thebibliography}{00}
\setlength{\itemsep}{-1pt}
\bibitem{iea2025} International Energy Agency, \textit{Electricity 2025: Analysis and Forecast to 2027}. Paris, France: IEA, 2025. [Online]. Available: https://www.iea.org/reports/electricity-2025

\bibitem{alkez2021}
D. A. Al Kez, A. M. Foley, F. W. Ahmed, M. O'Malley, and S. M. Muyeen,
``Potential of data centers for fast frequency response services in synchronously isolated power systems,''
\textit{Renewable and Sustainable Energy Reviews}, vol. 151, 2021, Art. no. 111547.




\bibitem{Fan2025}
Y. Fan, J. Zhao, and M. Yue, ``Adaptive Privacy Preserving Federated Learning for Virtual Power Plant Cyberattack Detection,'' \emph{IEEE Transactions on Smart Grid}, pp. 1--1, 2025, doi: 10.1109/TSG.2025.3646627.

\bibitem{dvorkin2025}
V. Dvorkin, ``Agent Coordination via Contextual Regression (AgentCONCUR) for Data Center Flexibility,'' \textit{IEEE Transactions on Power Systems}, vol. 40, no. 2, pp. 1832--1842, Mar. 2025.

\bibitem{ding2025}
Z. Ding \textit{et al.}, ``Multi-Time-Scale Joint Optimization of Data Center Market Transactions and Computing Resource Allocation,'' \textit{IEEE Transactions on Industry Applications}, 2025.

\bibitem{zou2025}
B. Zou, G. Chen, H. Zhang, and Y. Song, ``Coordinating Multiple Geo-Distributed Data Centers for Enhanced Participation in Frequency Regulation Services Under Uncertainty,'' \textit{Journal of Modern Power Systems and Clean Energy}, vol. 13, no. 5, pp. 1677--1688, Sep. 2025.

\bibitem{ren2025}
P. Ren, W. Sun, Y. Wang, and G. Harrison, ``Grid Frequency Stability Support Potential of Data Center: A Quantitative Assessment of Flexibility,'' arXiv preprint arXiv:2510.01050v1, Oct. 2025. [Online]. Available: https://arxiv.org/abs/2510.01050v1


\bibitem{wang2019} W. Wang, A. Abdolrashidi, N. Yu, and D. Wong, ``Frequency regulation service provision in data center with computational flexibility,'' \textit{Applied Energy}, vol. 251, 2019, Art. no. 113304. [Online]. Available: https://doi.org/10.1016/j.apenergy.2019.05.107


\bibitem{menati2023}
A. Menati, K. Lee, and L. X. Xie, ``Modeling and Analysis Utilizing Cryptocurrency Mining for Demand Flexibility in Electric Energy Systems A Synthetic Texas Grid Case Study,'' \textit{IEEE Transactions on Energy Markets, Policy and Regulation}, vol. 1, no. 1, pp. 1--10, Mar. 2023.

\bibitem{devries2023}
A. de Vries, ``The growing energy footprint of artificial intelligence,'' \textit{Joule}, vol. 7, no. 10, pp. 2191--2194, 2023. doi: 10.1016/j.joule.2023.09.004.


\bibitem{alibaba_clusterdata}
Alibaba Group, “Alibaba ClusterData: cluster-trace datasets from production data-centers,” GitHub repository alibaba/clusterdata, 2025. [Online]. Available: https://github.com/alibaba/clusterdata

\bibitem{iso-ne_markets}
ISO New England, “Markets \& Operations: Markets Data and Information,” 2025. [Online]. Available: https://www.iso-ne.com/markets-operations/markets

\bibitem{nyiso_data}
NYISO, “Energy Market \& Operational Data,” [Online]. Available: https://www.nyiso.com/energy-market-operational-data. Accessed: Nov. 5, 2025.

\end{thebibliography}
\end{document}